\documentclass[osajnl,twocolumn,showpacs,superscriptaddress,11pt]{revtex4-1}

\usepackage{epsfig} 
\usepackage{color} 
\usepackage{mathrsfs}
\usepackage{amsmath}
\usepackage{amssymb}
\usepackage{amsfonts}
\usepackage{mathrsfs}
\usepackage{color}
\usepackage{tikz}
\usepackage{epstopdf}

\newcommand{\imag}[1]{\text{Im}\left\{#1\right\}}
\newcommand{\real}[1]{\text{Re}\left\{#1\right\}}
\newcommand{\sign}{\text{sign}}

\begin{document} 


\newcommand{\mnras}{ Mon. Not. R. Astron. Soc.}
\newcommand{\amp}{\&}
\newcommand{\pasp}{Publ. Astron. Soc. Pac.}
\newcommand{\aeta}{Astron. Astrophys.}
\newcommand{\aap}{\aeta} 
\newcommand{\aaps}{Astronomy and Astrophysics Supplement}
\newcommand{\optcom}{Optics Communications}
\newcommand{\aplopt}{Applied Optics}
\newcommand{\optlet}{Optics Letters}
\newcommand{\solphys}{Solar Physics}

\newcommand{\ft}[1]{\mathscr{F}\left\{#1\right\}}
\newcommand{\ift}[1]{\mathscr{F}^{-1}\left\{#1\right\}}
\newcommand{\ftml}[1]{\mathscr{F}\big\{#1\big\}}
\newcommand{\npd}{M}
\newcommand{\nwfs}{M'}
\newcommand{\conv}{\ast}
\newcommand{\half}{0.5}
\newcommand{\quar}{0.25}


\newcommand{\xx}{\alpha}
\newcommand{\yy}{\beta}
\newcommand{\xda}{\alpha_{d1}}
\newcommand{\yda}{\beta_{d1}}
\newcommand{\xdb}{\alpha_{d2}}
\newcommand{\ydb}{\beta_{d2}}
\newcommand{\vrda}{\real{\ft{A_e \phi_{d1}}}}
\newcommand{\vida}{\imag{\ft{A_o \phi_{d1}}}}
\newcommand{\yrda}{\imag{\ft{A_e \phi_{d1}}}}
\newcommand{\yida}{\real{\ft{A_o \phi_{d1}}}}
\newcommand{\vrdb}{\real{\ft{A_e \phi_{d2}}}}
\newcommand{\vidb}{\imag{\ft{A_o \phi_{d2}}}}
\newcommand{\yrdb}{\imag{\ft{A_e \phi_{d2}}}}
\newcommand{\yidb}{\real{\ft{A_o \phi_{d2}}}}
\newcommand{\wfc}{\theta}
\newcommand{\pupipl}{E}
\newcommand{\focapl}{e}
\newcommand{\snd}{\xi}
\newcommand{\reqsupport}{N_\text{arr}}
\newcommand{\pupdim}{N_\text{pup}}
\newcommand{\regpa}{\epsilon}
\newcommand{\vest}{v_s}

\title{Fast \& Furious focal-plane wavefront sensing}

\author{Visa Korkiakoski} 
\affiliation{Delft Center for Systems and Control, Mekelweg 2, 2628CD Delft, The Netherlands}
\email{v.a.korkiakoski@tudelft.nl}
\affiliation{Leiden Observatory, Leiden University, P.O. Box 9513, 2300 RA Leiden, The Netherlands}

\author{Christoph U. Keller}
\affiliation{Leiden Observatory, Leiden University, P.O. Box 9513, 2300 RA Leiden, The Netherlands}

\author{Niek Doelman}
\affiliation{TNO Science and Industry, Stieltjesweg 1, 2628CK Delft, The Netherlands}
\affiliation{Leiden Observatory, Leiden University, P.O. Box 9513, 2300 RA Leiden, The Netherlands}

\author{Matthew Kenworthy}
\affiliation{Leiden Observatory, Leiden University, P.O. Box 9513, 2300 RA Leiden, The Netherlands}

\author{Gilles Otten}
\affiliation{Leiden Observatory, Leiden University, P.O. Box 9513, 2300 RA Leiden, The Netherlands}

\author{Michel Verhaegen}
\affiliation{Delft Center for Systems and Control, Mekelweg 2, 2628CD Delft, The Netherlands}



\begin{abstract}
We present two complementary algorithms suitable for using 
focal-plane measurements to control a wavefront corrector with an 
extremely high spatial resolution. The algorithms use linear
approximations to iteratively minimize the
aberrations seen by the focal-plane camera.
The first algorithm, Fast \& Furious (FF), uses a weak-aberration 
assumption and pupil symmetries to achieve fast wavefront reconstruction.
The second algorithm, an extension to FF, can deal with an arbitrary pupil
shape; it uses a Gerchberg-Saxton style error reduction
to determine the pupil amplitudes.
Simulations and experimental results are shown for 
a spatial light modulator controlling the wavefront with a resolution
of $170\times 170$ pixels. The algorithms increase the Strehl ratio
from $\sim$0.75 to 0.98--0.99, and the intensity of the scattered light is 
reduced throughout the whole recorded image of $320\times 320$ pixels. 
The remaining wavefront rms error is estimated to be $\sim$0.15 rad with FF and 
$\sim 0.10$ rad with FF-GS.
\end{abstract}

\maketitle

\section{Introduction}

When an object is imaged, variations of the refractive index in the medium, as well as optical alignment and manufacturing errors, distort the recorded image. This problem is typically solved using active or adaptive optics, where a deformable mirror, spatial light modulator (SLM) or a comparable device corrects the propagating wavefront. Typically, such systems are built with a separate optical arm to measure the distorted wavefront, because extracting the wavefront information from only focal-plane images is not trivial. However, focal-plane wavefront sensing is an active topic -- not only to simplify the optical design but also to eliminate the non-common path aberrations limiting the performance of high-contrast adaptive optics systems.

The most popular method for the focal-plane wavefront sensing is perhaps the Gerchberg-Saxton (GS) error reduction algorithm \cite{gerchberg1972, fienup82} and their variations, for instance \cite{green2003, burruss2010}. These are numerically very efficient algorithms, and it is easy to modify them for different applications. However, they suffer from accuracy, in particular because their iterative improvement procedure often stagnates at a local minimum. 

Various alternatives have been proposed, and a popular approach is to use general numerical optimization techniques  to minimize an error function;  examples include \cite{sauvage2007, riaud2012, paul2013}. However, when the number of optimization parameters is increased, the computational requirements generally rise unacceptably fast. The high computational costs are problematic for instance in astronomy; the largest future adaptive optics system is envisioned to have a wavefront corrector of a size of $200\times 200$ \cite{verinaud2010}. 

The numerical issues can be significantly reduced, if the unknown wavefront is sufficiently small. This is the case, for example, when calibrating the non-common path aberrations. Previous works have exploited the small-phase approximations \cite{giveon2007oe, meimon2010, martinache2013, smith2013}, but the implementations are generally not easily extended to the wavefront correction at extremely large resolution, such as over $100\times 100$ elements.

In this paper, we present two algorithms capable of extremely fast control of a wavefront correcting device with 20~000--30~000 degrees of freedom.

The first algorithm, Fast \& Furious (FF), has been published before \cite{keller2012spie,korkiakoski2012spie1,korkiakoski2013}.  It relies on small WF aberrations, pupil symmetries and phase-diversity to achieve very fast WF reconstruction. However, FF approximates the pupil amplitudes as an even function that not necessarily matches exactly the real situation. 

To improve the WF correction beyond the accuracy of FF, a natural way is to use approaches similar to the GS algorithm. However, the standard modifications of the algorithm are sensitive to the used phase diversities, in particular when the pupil amplitudes are not known, and they do not work with iterative wavefront correction as in FF. Therefore, our second algorithm combines FF and GS in a way that can be used not only to correct the wavefront, but also to estimate the pupil amplitudes -- for which we make no assumptions. This comes at a cost in terms of noise sensitivity and instabilities as well as more demanding computational requirements.

At first, we illustrate the motivation and principles of the FF
algorithm in Section~\ref{sec:ff}. Then, Section~\ref{sec:ffgs}
describes the Fast \& Furious Gerchberg-Saxton (FF-GS) algorithm in
detail. Section~\ref{sec:hardware} describes the used hardware,
Section~\ref{sec:results} shows simulation and experimental results,
and Section~\ref{sec:conclusions} draws the conclusions.

\section{Fast \& Furious}
\label{sec:ff}

\begin{figure*}[hbtp]  \center
\includegraphics[width=\textwidth]{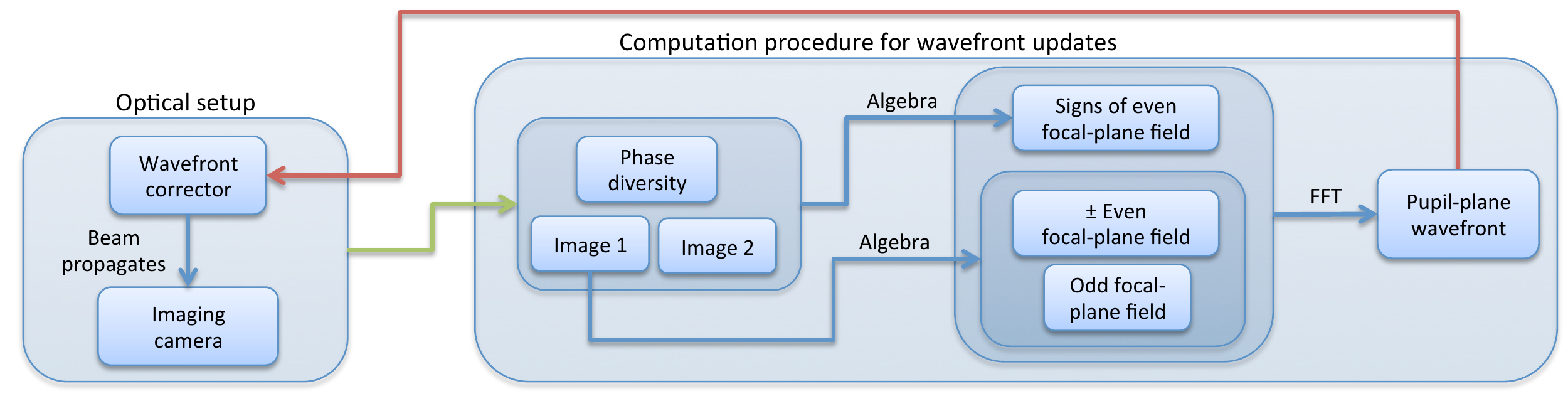} 
\caption{Schematic illustration of the FF algorithm.}
\label{fg:algoscemaff}
\end{figure*}

The Fast \& Furious algorithm is based on iteratively applying a
weak-phase approximation of the wavefront. The main principle
of the weak-phase solution is presented in  \cite{gonsalves2001}, 
but we found slight modifications \cite{keller2012spie}
leading to significantly better performance.
The algorithm uses focal-plane images and 
phase-diversity information to solve the wavefront, and the
estimated wavefront is corrected with a wavefront correcting
device. The correction step produces phase-diversity
information and a new image 
that are again used to compute the following phase
update. The schematic illustration of the algorithm is shown in 
Fig.~\ref{fg:algoscemaff}.

An important aspect of the algorithm is to maximize the use of the
most recent PSF -- denoted as Image 1 in Fig.~\ref{fg:algoscemaff}.
In the weak-phase regime, a single image is sufficient to estimate
both the full odd wavefront component and the modulus of the even
component of the focal-plane electric field. The phase-diversity is
needed only for the sign determination since we assume the wavefront aberrations are small. This makes the FF
substantially less prone to noise and stability issues as compared
to approaches relying more on the phase diversity 
information -- such as the FF-GS.

Section~\ref{sec:ffdet} explains the details of the weak-phase
solution, and Section~\ref{sec:ffpractical} discusses the practical
aspects when implementing the algorithm.

\subsection{Weak-phase solution}
\label{sec:ffdet}

A monochromatic PSF can be described by Fraunhofer diffraction
and is given by the squared modulus of the Fourier transform of the 
complex electric field in the pupil plane,
\begin{equation} \label{eq:pr}
  p = |\ft{A \exp(i\phi)}|^2,
\end{equation}
where $A$ is the pupil amplitude describing transmission and $\phi$
is the wavefront in the pupil plane.

The second order approximation of the PSF, in terms of the wavefront
expansion, can be written as
\begin{equation} \label{eq:p}
p = |\ft{A + iA\phi - 0.5A\phi^2}|^2.
\end{equation}
The phase $\phi$ can be represented as a sum of even and odd functions,
\begin{equation}
\phi = \phi_e + \phi_o,
\end{equation}
and Eq.~\eqref{eq:p} can then be written as
\begin{multline} \label{eq:p2}
p =  |\ftml{A + iA\phi_e + iA\phi_o \\
  - 0.5A\phi^2_e - 0.5A\phi^2_o - A\phi_e\phi_o}|^2.
\end{multline}
We make the assumption that $A$ is even, and therefore all the terms
here are either even or odd.  Therefore, the corresponding Fourier transforms
are then either purely real or imaginary with the same symmetries; we list
the corresponding terms in Table~\ref{tb:sym}.


\begin{table}[hbtp] \begin{center}
\caption{Notations and symmetries}
\label{tb:sym}
\begin{tabular}{lccclcc}
  \hline
  \multicolumn{3}{c}{Aperture plane}  & & \multicolumn{3}{c}{Fourier plane}  \\
  Term & Re/Im & Symmetry & & Term & Re/Im & Symmetry \\
  \hline
  $A$    & real & even   && a       & real & even \\
  $A\phi_e$ & real & even &&  $v$  & real & even\\
  $A\phi_o$ & real & odd   && $ iy$   & imaginary & odd\\
  $A\phi_e^2$ & real & even  &&  $v_2$  & real &  even\\
  $A\phi_o^2$ & real & even  && $y_2$  & real & even\\
  $A\phi_e \phi_o$ & real & odd  &&  $iz$ & imaginary & odd \\
  \hline
\end{tabular}\\
\end{center} \end{table}

Thus, all the introduced variables in Table~\ref{tb:sym} are purely real.
The quantities $a$, $v$ and $y$ denote the Fourier transforms of the
pupil function, even and odd wavefront aberrations, respectively, 
\begin{align}
a   &=  \ft{A}  \label{eq:a} \\
v   &=  \ft{A\phi_e}  \label{eq:v} \\
y   &= \imag{ \ft{A\phi_o} } \label{eq:y}.
\end{align}
Using the definitions, the second-order PSF approximation can be written as
\begin{equation}
p = |a + iv - y - \half v_2 - \half y_2 -i z|^2,
\end{equation}
which simplifies to 
\begin{equation}
  p = a^2 + v^2 + y^2 - 2ay + \snd,
\end{equation}
where the first four terms constitute the first order approximation -- in
terms of the wavefront expansion -- and the second-order component is
\begin{multline} \label{eq:snd}
  \snd = \quar v_2^2 +\quar y_2^2 +z^2 -av_2 -ay_2 +\half v_2 y_2 \\
   + yv_2 +yy_2 - 2vz.
\end{multline}

\begin{figure}[hbtp]  \center
\includegraphics{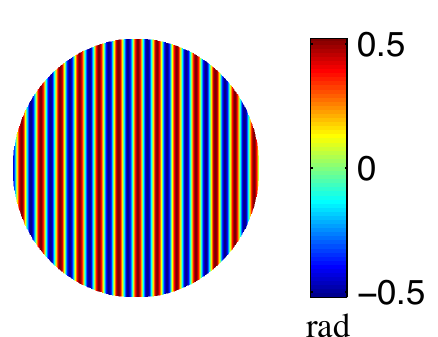} 
\includegraphics{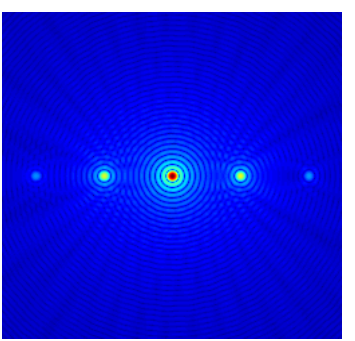} 
\caption{Left: a purely sinusoidal wavefront. Right: resulting image raised
  to the power of 0.2 to compress the dynamic range.}
\label{fg:wfsample}
\end{figure}

The above equations are best illustrated by an example.  We consider a
purely sinusoidal wavefront having a peak-to-valley value of 1.0 rad and
an rms error of 0.37 rad -- alternative examples can be seen for instance 
in \cite{perrin2003}. The wavefront and the resulting PSF image
are shown in Fig.~\ref{fg:wfsample}.  The WF causes two main
side lobes and more side lobes with significantly lower intensity; one
pair is shown in Fig.~\ref{fg:wfsample}.

\begin{figure}[hbtp]  \center
\includegraphics{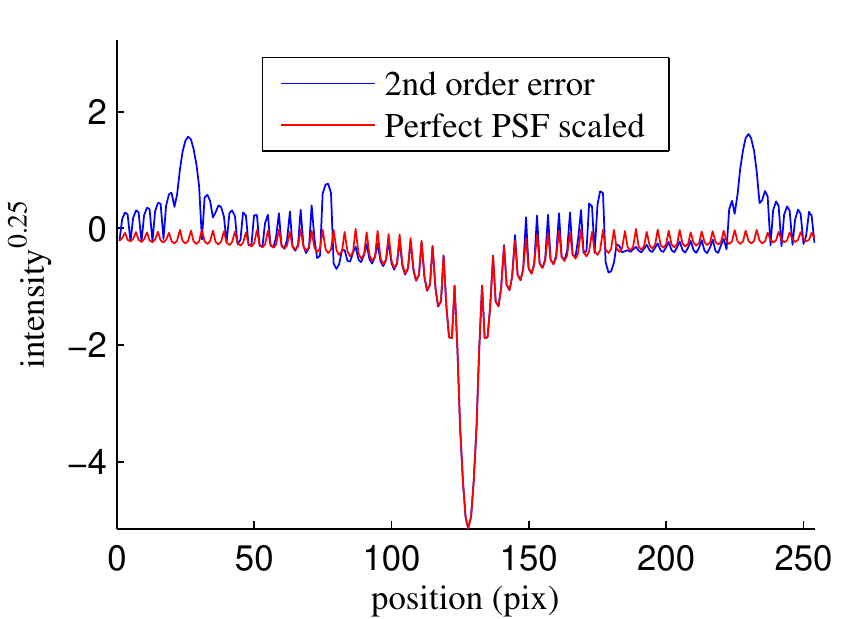} 
\caption{Radial cuts of the second order component $\snd$, defined in
  Eq.~\eqref{eq:snd}, and an inverted and scaled perfect PSF, $a^2$.}
\label{fg:ffradcuts1}
\end{figure}

Fig.~\ref{fg:ffradcuts1} shows a radial cut of the second order component $\snd$ for the example wavefront. Its most significant terms are $av_2$ and  $ay_2$, and therefore the perfect image ($a^2$) scaled by a negative coefficient approximates $\snd$ reasonably well. This term is responsible of the energy conservation by reducing the Strehl ratio \cite{keller2012spie}. The first-order approximation always has a Strehl ratio of 1.



\begin{figure}[hbtp]  \center
\includegraphics{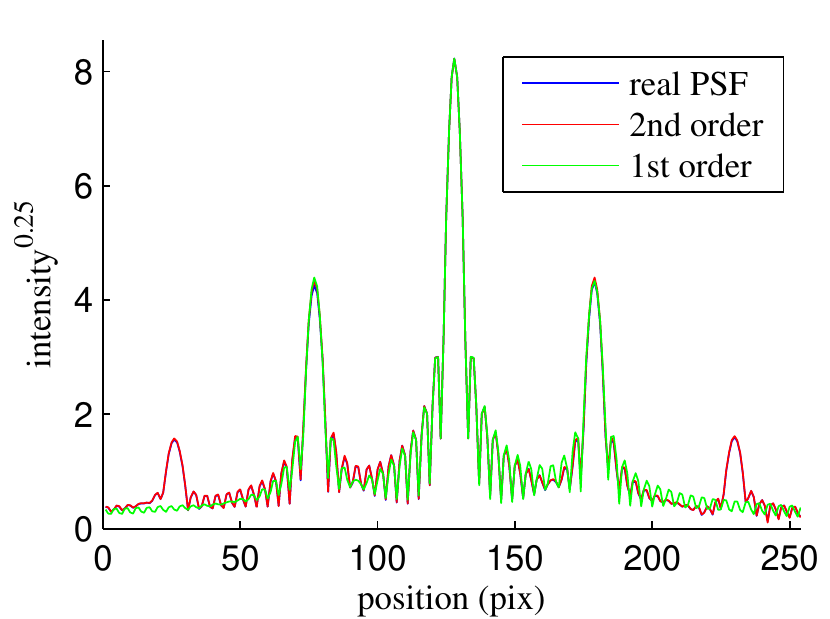} 
\caption{Radial cuts of the perfect PSF, its improved 1st order 
  approximation and the 2nd order approximation. The latter is
  virtually identical to the perfect PSF.}
\label{fg:ffradcuts2}
\end{figure}

Thus, an improved first order approximation can be obtained by
subtracting a scaled version of $a^2$ from the first order PSF
approximation; the scaling coefficient needs to be adjusted such that the
maxima of the perfect PSF and the approximation are the same. The radial
cuts of the PSF approximations are illustrated in
Fig.~\ref{fg:ffradcuts2}. The improved first-order approximation captures
the main lobe and the first pair of side lobes quite well, but the secondary
side lobes are missed.

However, for a wavefront with an rms error of less than one radian, the
improved first-order approximation is often sufficient, and it can be 
formulated as
\begin{equation} \label{eq:p1}
  p = a^2+y^2+v^2 -2ay
  -\left(1-\frac{\max{\left(p_n\right)}}{\max\left(a^2\right)}\right)a^2,
\end{equation}
where $p_n$ denotes the recorded image normalized to the same
energy as the perfect PSF,
\begin{equation} \label{eq:pnorm}
  p_n = p_m \frac{\sum_{x,y} a^2(x,y)}{\sum_{x,y} p_m(x,y)},
\end{equation}
where $(x,y)$ denotes the detector pixel coordinates and $p_m$ 
is the raw image. Therefore, to simplify the notations, it is convenient to
define a modified normalization of a PSF,
\begin{equation} \label{eq:scaling}
  p' =p_n+\left(1-\frac{\max(p_n)}{\max\left(a^2\right)}\right)a^2,
\end{equation}
where the normalized image, $p'$, has the same maximum as $a^2$.

To solve the wavefront using Eq.~\eqref{eq:p1}, we follow the 
procedure of \cite{gonsalves2001}, which is repeated
here for convenience. 

The recorded image is normalized and broken to its even and odd
parts. It then holds that
\begin{align}
  p'_e &= a^2 +v^2 +y^2 \label{eq:pe} \\
  p'_o &= 2ay. \label{eq:po}
\end{align}
The odd component of the wavefront is then easily reconstructed by
first solving $y$ using Eq.~\eqref{eq:po}, and then using the
inverse of Eq.~\eqref{eq:y}.  Due to noise and approximation 
errors, the direct application of
Eq.~\eqref{eq:po}, however, would result in division by excessively
small values. We compensate this by using a regularization 
as in \cite{gonsalves2001},
\begin{equation} \label{eq:yreg}
  y = \frac{a p'_o}{2a^2 + \regpa},
\end{equation}
where $\regpa $ is a small number. We found it best to set $\regpa$
to a value of 50--500 times the measured noise level of the recorded images.

To compute the even wavefront component, we need additional
information in the form of phase diversity. We assume that a
second, previously recorded image is known, and it was obtained
with a known phase change compared to $p$. The even component of 
its normalized version can be written as
\begin{equation} \label{eq:pe2}
  p'_{e2} = a^2 +(v+v_d)^2 +(y+y_d)^2,
\end{equation}
where $v_d$ and $y_d$ are the even and odd Fourier components of the
phase diversity, obtained in analogy to Eqs.~\eqref{eq:v} and
\eqref{eq:y}.

Using Eqs.~\eqref{eq:pe} and \eqref{eq:pe2}, we can solve $v$ (the even phase component in Fourier space) and write it as
\begin{equation} \label{eq:vs}
  \vest = \frac{p'_e - p'_{e2} -v_d^2 -y_d^2 -2yy_d}{2v_d}.
\end{equation}
However, this formula is highly sensitive to noise due to the
subtraction of two very similar images. Therefore, as also in
\cite{gonsalves2001}, we use Eq.~\eqref{eq:vs} only to compute the signs of
$v$; a more robust form follows from the use of Eq.~\eqref{eq:pe},
\begin{equation} \label{eq:vsolv}
  v = \sign\left(\vest \right) \left|p'_e - a^2 - y^2\right|^{0.5},
\end{equation}
where we use the absolute value to avoid taking the square root of negative
values, occurring due to noise and approximation errors; this was
observed to work better than zeroing the negative values. The even
wavefront component is then computed in the same way as the odd one,
by using Eq.~\eqref{eq:vsolv} and the inverse of Eq.~\eqref{eq:v}.

\subsection{Practical aspects}
\label{sec:ffpractical}

To use the FF algorithm as presented here, it is necessary to have a wavefront correcting device -- a deformable mirror or spatial light modulator -- whose phase response is known. It is then possible to translate the desired phase change to appropriate wavefront corrector command signals. An appropriate mapping can be created using the standard adaptive optics calibration procedures as in \cite{korkiakoski2012spie1} or, as we do here, with the help of dOTF based calibration method \cite{korkiakoski2013}.  The method is based on determining the SLM phase (and transmission) response, when the control signal is changed in different pixel blocks. This data is then used to find an affine transform that maps the location of each SLM pixel to its physical location in the pupil plane.

We also assume that the collected images are sufficiently sampled: without aberrations the full width at half maximum of the PSF has to be at least two pixels.  If the detector is undersampled, aliasing prevents using the intensity images as described in Section~\ref{sec:ffdet}. Large oversampling is also not desired since it increases the computational requirements.

The phase array, $\phi$, needs to be sampled with sufficient resolution
to also model the pupil aperture, $A$, with good accuracy. The values we use
($170\times 170$) are sufficient for our purpose; we expect no significant
sampling errors when implementing Eqs.~\eqref{eq:v} and \eqref{eq:y} as 
fast Fourier transforms (FFTs). 
However, we need to zero-pad the recorded images such that
the FFTs correctly implement the Fourier transforms in 
Eqs.~\eqref{eq:a}, \eqref{eq:v} and \eqref{eq:y}; the sampling of the arrays $a$,
$v$ and $y$ need to match the pixels of the  camera. The 
amount of zero-padding is determined by the sampling 
coefficient,
\begin{equation} \label{eq:q}
  q = \frac{\reqsupport}{\pupdim},
\end{equation}
where $\reqsupport$ is the dimension of the FFT array and $\pupdim$
is the size of $\phi$. We use the dOTF method as discussed in \cite{korkiakoski2013} to find $q$. The method is based on the use of localized phase diversity at the pupil border, which makes it possible to very straightforwardly create an array where the pupil shape can be directly seen. The parameter $q$ is calculated by comparing the sizes of the pupil and the dOTF array.

When performing the FFT to obtain the phase from $v$ and $y$, we combine
the two real-valued FFTs to a single complex FFT \cite{keller2012spie},
\begin{equation} \label{eq:phisol}
  A\phi = \ift{w\left(v + iy\right)},
\end{equation}
where $w$ is a windowing function; it implements filtering
necessary for numerical regularization -- typically, high spatial frequencies
are detected with higher uncertainty, and they need to be damped to obtain
feasible reconstructions. The regularization is needed also with noiseless 
images since the weak-phase solution 
provides only approximate wavefronts. In this work, we have used a concave
parabola, whose width can be adjusted depending on the noise level.
An optimum filter is the subject of future studies.

To implement the iterative feed-back loop to optimize the wavefront
error, we use a standard, leaky-integrator control. The
wavefront-corrector shape at time step $k$ is calculated as
\begin{equation} \label{eq:feedback}
  \theta_k = g_l\theta_{k-1} -gA\phi_{k-1},
\end{equation}
where $g_l$ is the leaky gain, $\theta_{k-1}$ is the previous wavefront
corrector shape, $g$ is the integrator gain, and
$A\phi_{k-1}$ is the most recent small phase solution, computed using
two most recent images using Eq.~\eqref{eq:phisol}.

The integrator gain, $g$, determines the tradeoff between convergence
speed and stability; a small gain results in slow convergence, a high
gain means the image noise causes larger errors after the algorithm
has converged. Excessively small gain would also make the use
of phase-diversity information difficult.

The leaky gain is another regularization parameter. 
A value of $g_l=1$ would be equal to a standard
integrator, and it would be optimal in the case of no errors, the
equation $p=|\ft{A\exp(i\phi)}|^2$ perfectly describing the
system. Values $g_l < 1$ introduce wavefront aberrations
at every time step preventing the system reaching a perfect state.
However, that also prevents creeping instabilities from
destroying the performance. The result is a stable convergence to a
level with a slightly higher residual wavefront error.


\section{Fast \& Furious Gerchberg-Saxton}
\label{sec:ffgs}

The obvious limitation of the FF algorithm is the assumption of the pupil
amplitudes being even. This holds reasonably well for most
of the optical systems having a circular shape, possibly with a
central obstruction. However, to achieve the optimal focal-plane
wavefront sensing with a high-order system not suffering from other
limiting factors, it is necessary to consider imaging models where the
pupil amplitudes can have an arbitrary shape.

We have approached the problem by combining the FF-style weak-phase
solution and a version of the Gerchberg-Saxton (GS) algorithm. The new
algorithm is referred to as FF-GS in the following.

\begin{figure*}[hbtp]  \center
\includegraphics[width=\textwidth]{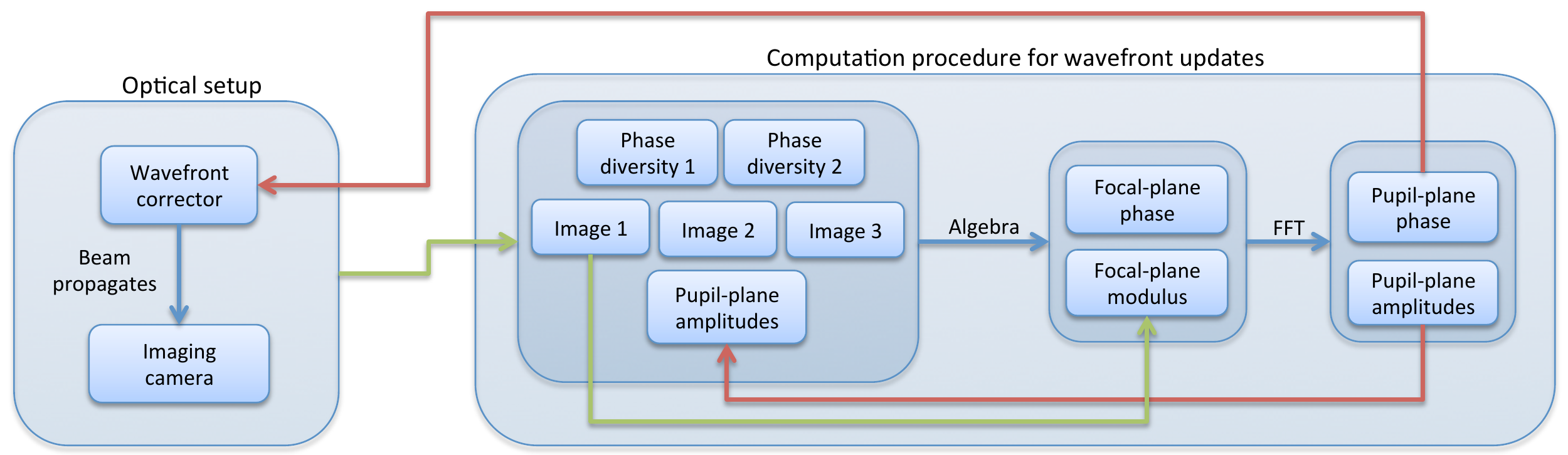} 
\caption{Schematic illustration of the FF-GS algorithm.}
\label{fg:algoscema}
\end{figure*}

As with the GS algorithm, we maintain an iteratively updated estimate
of the unknown quantities -- in our case the pupil amplitudes. The
pupil amplitude estimate, phase diversities and the recorded images
are used to calculate the focal-plane field; it requires three
Fourier transforms and the use of the weak-phase approximation. Then,
a Fourier transform is used to propagate the field to the pupil
plane. The propagation results in improved estimates for the
pupil-plane amplitudes and the wavefront.
The schematic illustration of the FF-GS algorithm is shown in
Fig.~\ref{fg:algoscema}.

The FF-GS computation procedure forms a loop that could
be iterated several times to obtain improved wavefront estimates.
However, we found that in practice it is sufficient to run only two
iterations before applying the wavefront correction with 
the obtained estimate. As with FF, the wavefront correction
yields another image and phase-diversity information, which
are used to compute the following correction step.

Next, Section~\ref{sec:ffgsdet} describes the algebra that we
use to compute the focal-plane electric field during the FF-GS
procedure. Then, Section~\ref{sec:ffgsiters} explains the details of
the iterative computation, and Section~\ref{sec:ffgspractical}
discusses practical issues we face when implementing the algorithm.

\subsection{A more general weak-phase solution}
\label{sec:ffgsdet}

In this section, we assume that an approximation of the pupil amplitudes
(denoted here as $A$) is known; as a first step, a top-hat function
is sufficient in the case of an unobstructed, round pupil. 
The estimates are updated iteratively,
and we will make no restrictive assumptions about $A$.

We assume that three images are collected and that the corresponding
phase-diversity information is known. The images are normalized
according to Eq.~\eqref{eq:scaling}, and it holds approximately that
\begin{align} 
  p'_1&=|\focapl_1|^2= |\ft{A + iA\left(\phi\right)}|^2 \label{eq:ffgs1} \\
  p'_2&=|\focapl_2|^2=|\ft{A+iA\left(\phi+\phi_{d1}\right)}|^2 \label{eq:ffgs2}\\
  p'_3&=|\focapl_3|^2=|\ft{A+iA\left(\phi+\phi_{d2}\right)}|^2, \label{eq:ffgs3}
\end{align}
where $\focapl_1$, $\focapl_2$ and $\focapl_3$ are the electric fields
corresponding to the images, $\phi$ is the unknown pupil-plane phase, 
and $\phi_{d1}$ and $\phi_{d2}$ are
the known phase diversities applied to successively recorded images.

When counting the number of unknown variables, one can see that it might be possible to solve the unknown phase using only two images, with Eqs.~\eqref{eq:ffgs1} and \eqref{eq:ffgs2}. However, we found the following procedure with three images to be better. In addition of making the algebra easier, it is also significantly more robust since more information is available to compensate the errors in the estimate of $A$. Using even more images could potentially still improve the results, but studying this is outside the scope of this paper.

Instead of solving the phase directly, we use phase-diversity
information to find the electric field at the focal-plane.  The
electric field corresponding to Eq.~\eqref{eq:ffgs1} can be written as
\begin{equation} \label{eq:elf}
  \focapl_1 =\left(a_r +\xx\right) +i\left(a_i+\yy\right),\\
\end{equation}
where
\begin{align}
  a_r &= \real{\ft{A}} \nonumber\\
  a_i &= \imag{\ft{A}} \nonumber\\
  \xx &= -\imag{\ft{A\phi}} \nonumber\\
  \yy &= \real{ \ft{A\phi}} \nonumber.
\end{align}
The unknown coefficients $\xx$ and $\yy$ can be found by solving the
equations that follow when subtracting Eq.~\eqref{eq:ffgs1} from
Eqs.~\eqref{eq:ffgs2} and \eqref{eq:ffgs3}. The subtraction cancels
all the non-linear terms and results in linear equations,
\begin{equation} \label{eq:focacoefs}
  \left[\begin{array}{cc}
      2\xda & 2\yda \\
      2\xdb & 2\ydb 
    \end{array} \right] 
  \left[\begin{array}{c}
      \xx \\ \yy \end{array} \right]
  =
  \left[\begin{array}{c}
      c_1 \\ c_2 \end{array} \right],
\end{equation}
where
\begin{align}
  \xda &= -\imag{\ft{A\phi_{d1}}} \nonumber \\
  \yda &=  \real{\ft{A\phi_{d1}}} \nonumber \\
  \xdb &= -\imag{\ft{A\phi_{d2}}} \nonumber \\
  \ydb &=  \real{\ft{A\phi_{d2}}}, \nonumber 
\end{align}
and
\begin{align} \label{eq:psubst}
  c_1 &= p'_2-p'_1 -\left(2a_r\xda +2a_i\yda +\xda^2+\yda^2\right)\nonumber \\
  c_2 &= p'_3-p'_1 -\left(2a_r\xdb +2a_i\ydb +\xdb^2+\ydb^2\right).
\end{align}
%


We solve the coefficients $\xx$ and $\yy$ by inverting the $2\times 2$ matrix in Eq.~\eqref{eq:focacoefs}. The matrix has full rank, if the used phase-diversities are linearly independent. We found this generally to be the case when applying the algorithm, and therefore it was unnecessary to use any regularization methods. The coefficients can then be substituted into Eq.~\eqref{eq:elf} to compute the focal plane electric field. However, this estimate would again be very prone to noise due to the subtraction of similar images, as shown in Eq.~\eqref{eq:psubst}. Therefore, it is better to use the directly measured modulus and use only the phase-information following from Eq.~\eqref{eq:elf}. This then gives a more robust focal-plane estimate,
\begin{equation} \label{eq:focaplfix}
  \focapl_1 =\left|p'_1\right|^{0.5}\exp\left[i\arg((a_r+\xx)+i(a_i+\yy))\right].
\end{equation}
The following section explains the details how this is then combined with 
the GS approach.

\subsection{Iterative computation procedure}
\label{sec:ffgsiters}

As the previous section indicates, we first record
three images. The phase-diversity can be chosen freely, as long as its 
peak-to-valley stays below one radian. We use the FF algorithm at the initial 
steps.

Then, using the collected data, we perform computations to calculate a
new wavefront update. The wavefront update
is applied, and another image with different phase-diversity
information is collected.  The three most recent images are then used again
to calculate the next phase correction to be applied. We continue
until the algorithm converges.

The computation consists of a cycle of two successive GS-like
iterations.  The complete process consists of the following steps:
\begin{enumerate}
\item Take the pupil amplitudes, $A$, estimated at the previous
  iteration. Use the procedure in Section~\ref{sec:ffgsdet} to
  calculate the focal-plane electric field corresponding to $p_2$,
  the second most recent image. This is be done by solving $\xx$ and $\yy$
  in Eq.~\eqref{eq:focacoefs}  and using formula
    \begin{displaymath}
    \focapl_2  = \left|p'_2\right|^{0.5}\exp\left[
      i\arg\left((a_r+\xx)+i(a_i+\yy)\right)\right].
  \end{displaymath}
  Here, the images could be rearranged appropriately: $p_2$
  should be the reference and the phase diversities interpreted accordingly.
  However, we found $\arg(\focapl_2)]\approx\arg(\focapl_1)$ to be a sufficient
  approximation.
  
\item Compute the pupil-plane electric field corresponding to the
  image $p_2$. This is done by Fourier transforming the focal-plane
  field,
  \begin{displaymath}
    \pupipl_2  = \ift{\focapl_2}.
  \end{displaymath}

\item Update the current estimate of the pupil amplitudes:
  \begin{displaymath}
    A = |\pupipl_2|.
  \end{displaymath}
  
\item With the new pupil amplitude estimate, repeat the procedure in
  Section~\ref{sec:ffgsdet} to compute the electric field for
  image $p_1$, the most recent image.

\item Compute the pupil-plane field corresponding to image $p_1$,
\begin{displaymath}
  \pupipl_1  = \ift{\focapl_1}.
\end{displaymath}

\item Calculate the final phase estimates for the phase and pupil
  amplitudes,
\begin{align}
  \phi &= \arg(\pupipl_1) \label{eq:phigs} \\
  A    &= |\pupipl_1|. \label{eq:ags}
\end{align}
\end{enumerate}


The estimates of $\phi$ are then used in the feedback loop 
in the same way as with the FF algorithm.

\subsection{Practical aspects}
\label{sec:ffgspractical}

The issues faced in practice by an implementation of the FF-GS differ slightly
from the simple FF. 

Since the pupil amplitudes are not constrained, the imaging model is
potentially much more accurate. In practice, indeed, we found that it
was not necessary to apply any windowing filters to dampen the high
spatial frequencies in the wavefronts reconstructed with FF-GS.
The normal feed-back loop, as described by Eq.~\eqref{eq:feedback}, was
sufficient regularization for the optimal performance.

It was also not necessary to introduce any ad-hoc restrictions to
constrain the pupil amplitudes. The values obtained from
Eq.~\eqref{eq:ags}, at any time step, do have a significant deviation
from the actual pupil amplitudes, but this appears to be a minor issue
for the convergence of the algorithm. Moreover, averaging the values
of $A$ over several iterations produces non-biased results.

However, the heavier reliance on the phase-diversity information makes
the algorithm more prone to stability issues.  To increase the
stability, we found it helpful to introduce other ad-hoc techniques.

In the feedback loop, we apply amplitude gains. Just as formulated in
Eq.~\eqref{eq:feedback}, we multiply the applied phase correction --
obtained from Eq.~\eqref{eq:phigs} -- by the estimated
amplitudes. This helps to prevent abrupt phase changes at points where
$|\pupipl_1|$ has a very small value; at those points, the
determination of the complex phase is likely to fail. In fact,
we also set $\phi$ to zero at points where $|\pupipl_1|<0.3$.  This
reduces the speed of convergence, but has no impact on the accuracy of
the converged solution.

Finally, additional regularization is used in case of numerical issues 
when the algorithm has converged. We observed that occasionally, every 10th
iteration or so, the FF-GS algorithm produces wildly incorrect
results. This is related to the fact that the solution of Eq.~\eqref{eq:focacoefs}
requires phase-diversity information. Once the applied phase corrections
become very small, the corresponding diversity information
becomes unreliable.

To make sure that such violent phase changes will not cause troubles,
we simply restrict the magnitude of the applied phase change. If the rms
value of the change exceeds the mean of ten previous changes, we scale
it down to the mean value.

\section{Hardware used}
\label{sec:hardware}

To test the algorithms, we created a simple setup that consists of one
spatial light modulator (SLM) and an imaging camera. The former is a
reflective device (BNS P512) having a screen of $512\times 512$~pixels,
a fill factor of 83.4\% and a pixel pitch of $15\times 15 \mu$m.
The SLM is able to create a phase-change of 2$\pi$~radians at the used
wavelength, and its control signal is coded with 6 bits.

The imaging camera is a Basler piA640-210gm, which has a resolution of
$648\times 488$~pixels and a dynamic range of 12 bits. As a light
source, we use a fibre-coupled laser diode (Qphotonics QFLD-660-2S)
having a wavelength of 656~nm.

\begin{figure}[hbtp]  \center
\includegraphics[width=\columnwidth]{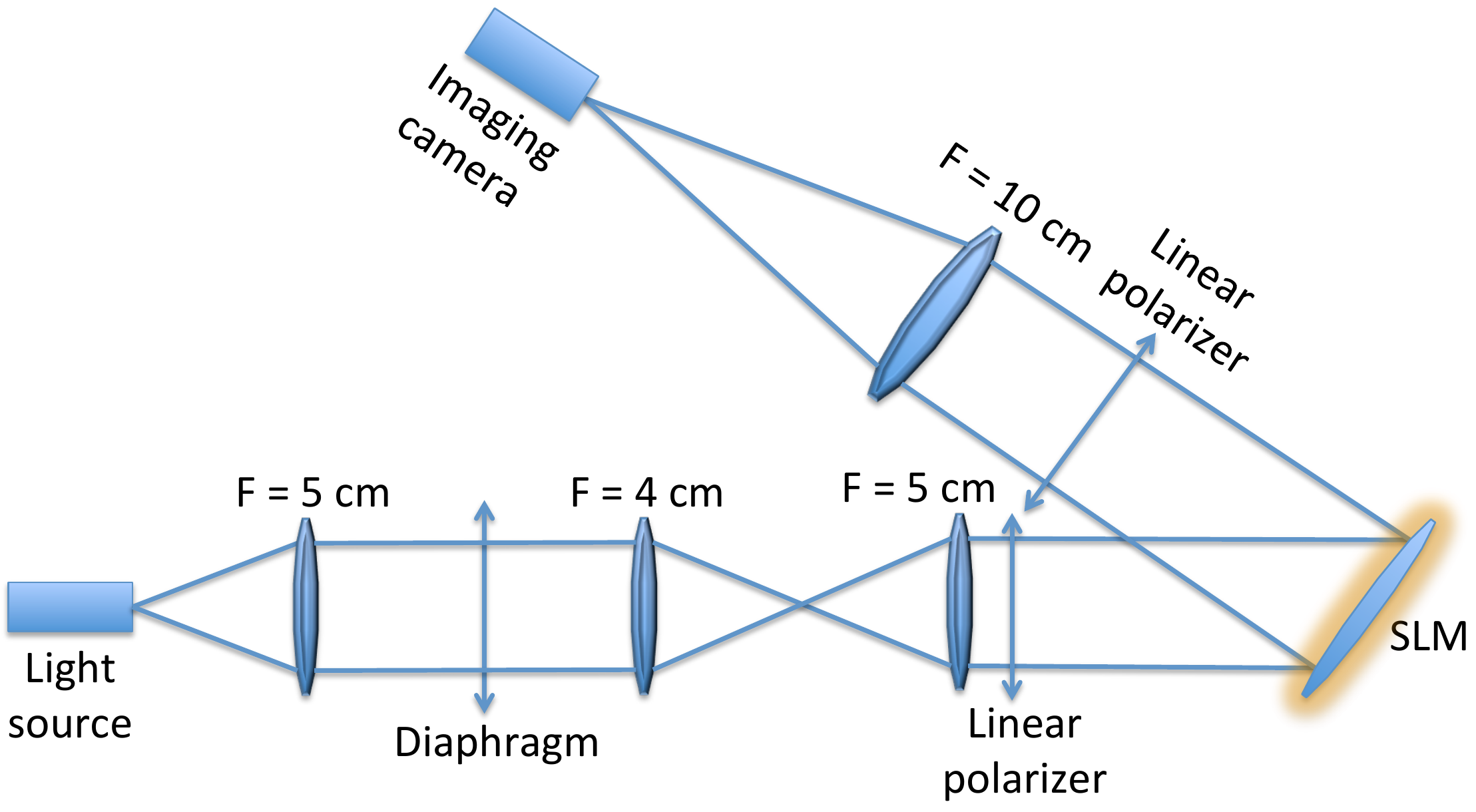} 
\caption{Schematic view of the used hardware. The lenses are standard 1-inch doublets. The beam diameter is 3.7~mm at the SLM.}
\label{fg:setupschema}
\end{figure}

A schematic figure of the setup is shown in Fig.~\ref{fg:setupschema}. The beam goes first through a diaphragm, and it is then collimated such
that it hits an area of $245\times 245$~pixels on the SLM. 
The device reflects several sub-beams due to strong diffraction effects,
and we use only the zeroth order beam; it is directly imaged onto the camera (beam numerical aperture NA=0.037). 
The other sub-beams 
cause no adverse effects.  Before and
after the SLM, we place two linear polarizers that are rotated such
that their orientation matches the one of the SLM. 

\begin{figure}[hbtp]  \center
\includegraphics[width=\columnwidth]{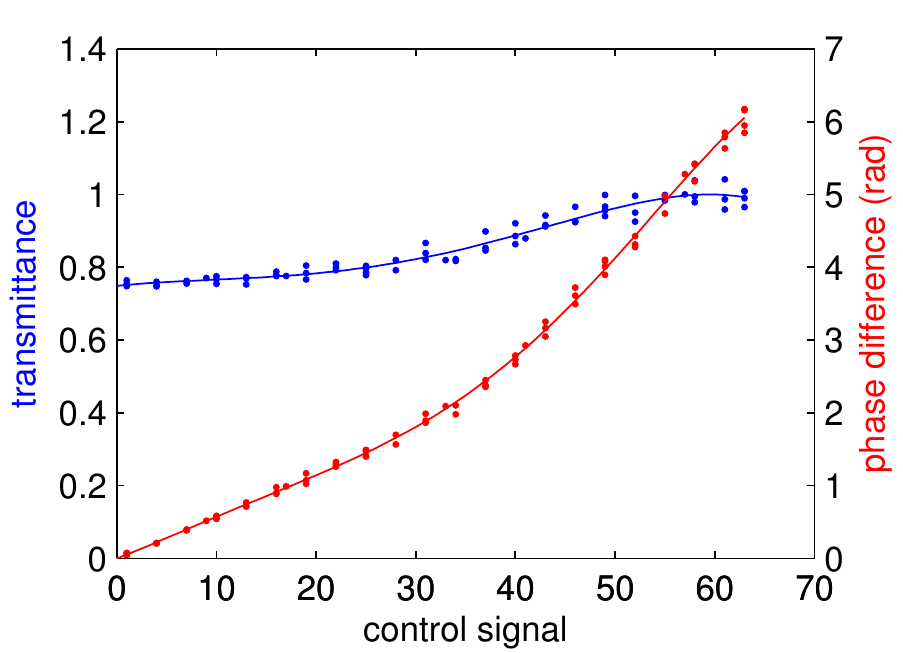} 
\caption{SLM phase and amplitude responses. The dots indicate individual
  measurements. The lines show 5th order polynomial fits to the data.}
\label{fg:slmrespo}
\end{figure}

The SLM phase and transmittance responses are measured with the
differential optical transfer function (dOTF) method as described in
\cite{korkiakoski2013}. The resulting measurements are shown in
Fig.~\ref{fg:slmrespo}. The maximum control voltage causes
$\sim$$2\pi$ phase shift at 656~nm.

The used SLM couples the transmittance and phase change; the
transmittance gradually increases when a larger phase shift is
introduced with the SLM. For phase changes of less than one radian, the
transmittance is $\sim$25\% lower compared to what is seen when a
change of more than $\sim$4~rad is introduced.

To create a mapping between the pupil-plane coordinates and the SLM
pixels, we again use the dOTF method and affine transforms as described
in \cite{korkiakoski2013}. This time, however, we make the dOTF
recording in the best focus to avoid issues with the non-telecentric
beam. To compensate for signal-to-noise problems, we take more images
to average out the noise: it takes $\sim$2~hours to create one dOTF
array. This makes the process also more vulnerable to internal
turbulence in the setup; the recorded images are blurred
such that the low spatial frequencies in the images become distorted,
and we have to mask out the center of the obtained dOTF arrays.

\begin{figure}[hbtp]  \center
\includegraphics[width=\columnwidth]{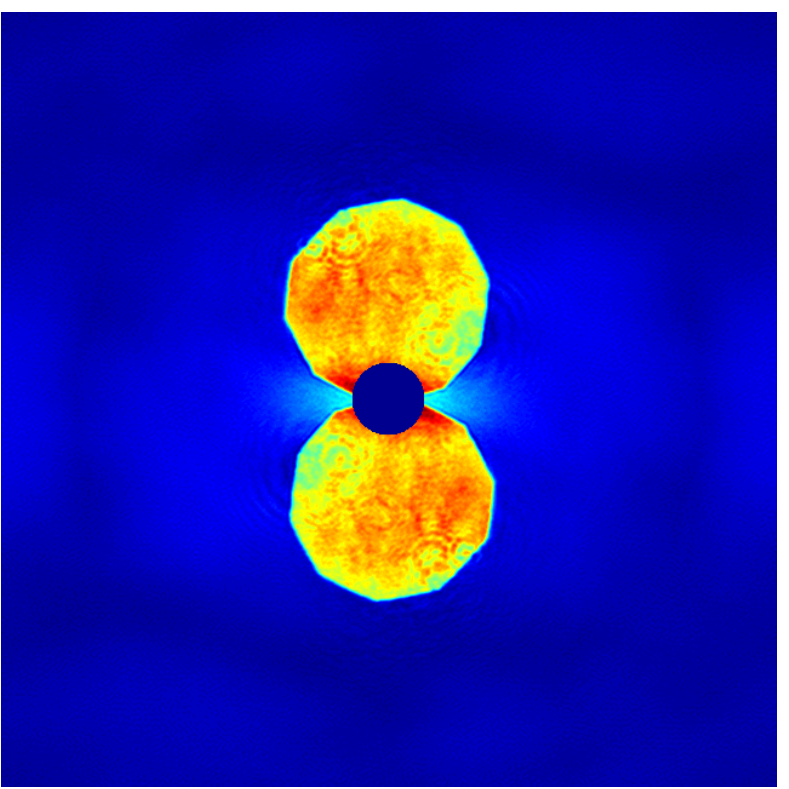}  
\caption{The modulus of an averaged dOTF array.}
\label{fg:dotf}
\end{figure}

Fig.~\ref{fg:dotf} shows the modulus of the best-focus
dOTF array recorded with the whole SLM at zero control voltage. 
Although the center of the array is masked, it is
still perfectly usable for the calibration
process of \cite{korkiakoski2013}, and we can accurately determine the
PSF sampling as defined by Eq.~\eqref{eq:q}: $q=3.76\pm0.01$.

The resulting SLM calibration is valid as long as the position of the
SLM stays fixed with respect to the imaging camera, and the
phase-response of the device does not change. In our setup, we found
this to be case for at least one month -- from the initial calibration to
the last measurements reported in this paper.

As discussed in \cite{korkiakoski2013}, the resolution of the
controlled phase is a free parameter when calculating the affine
mapping for the SLM calibration. We obtained good results when using
$\sim$30\% less pixels than are actually used by the SLM. Thus, we
selected the size of the controlled phase array as $\pupdim=170$. The
resulting FFT array dimension is then $\reqsupport=640$.

When recording images for the FF and FF-GS algorithms, we use the same high-dynamic range (HDR) imaging approach as in \cite{korkiakoski2013}.  Several snapshot images are taken with different exposure times, and we combine the images to extend the dynamic range and compensate noise. Each single-exposure component in one HDR image is an average over 40--200 images, and we used in total 16 exposure times (2, 5, 12, 25, 50, 100, 200, 400, 750, 1100, 1450, 1800, 2150, 2500, 2850 and 3200~ms). It took $\sim$15~s to record one HDR image.  Increasing the integration even further does not significantly improve the performance of the wavefront correction algorithms.

Although the imaging camera has a resolution of $640\times
480$~pixels, we use only a smaller area for convenience reasons. After
acquiring the image, we crop an array of $320\times 320$~pixels such
that the PSF maximum is in the center. Outside of the region, we did not
observe any significant amount of light.

To detect all the spatial frequencies corrected by the controlled
phase array of $170\times 170$~pixels, however, we would need an array
of $640\times 640$~pixels. Thus, it is possible that our control
algorithms introduce high-spatial frequencies that scatter light
outside of the observed image. However, with FF, this is mitigated by
the applied low-pass filter. With FF-GS, we observed no stability
issues with the high spatial frequencies, although no explicit
regularization measures were taken.

\section{Results}
\label{sec:results}

This section illustrates the results of the FF and FF-GS algorithms.
We consider only a single case: the wavefront to be corrected
is what the camera sees at the beginning, when no voltage is applied
to the SLM. We call this the initial situation.

We concentrate on the ultimate accuracy the algorithms can achieve in a low-noise regime. Our earlier publication  \cite{korkiakoski2012spie1} describes in more detail the FF performance in the presence of more noise. We showed that the algorithm works, but only the lower spatial frequencies can be reconstructed. Now, we study a case that is typical for a high-order adaptive optics test bench, and the noise level is chosen such that FF-GS offers an advantage over FF -- with higher noise FF is more robust.

Section~\ref{sec:study} illustrates the properties of
the converged algorithms as measured with our test
setup. Section~\ref{sec:meassimu} shows a more detailed comparison of
the measurements and simulations with the actual hardware modeled in
sufficient detail. Finally, Section~\ref{sec:errbud} presents a
simulation-based error budget that quantifies the effects of different
error sources.


\subsection{Performance of the algorithms}
\label{sec:study}

For the results shown here, we have optimized the free parameters
(FF regularization coefficient $\regpa$, the width of the FF 
filtering window $w$, leaky gain $g_l$, loop gain $g$) such
that the converged WF quality is best; the convergence speed has lower priority.

The width of the filtering window used by the FF algorithm was chosen
to be $320\times 320$, the same as the recorded images. However, during the first 10
iterations, we used a narrower window (width of 80~pixels) to avoid
introducing errors at the high spatial frequencies. After the lower
spatial frequencies are corrected, it is safe to increase the window
size. 

The optimal values for feedback loop gains were $g=0.3$, $g_l=0.97$
(with FF) or $g_l=0.999$ (with FF-GS), and $\regpa$ was 250 times the
determined noise level in the images.

For the FF algorithm, we also need to determine the pupil amplitudes,
$A$.  We use a perfect top-hat function having a size of $\pupdim
\times \pupdim$, where the choice of $\pupdim$ is explained in
Section~\ref{sec:hardware}. It might be possible to improve the
results by adjusting $A$ based on the actual pupil shape, but this is
outside the scope of this paper.

With these settings, both FF and FF-GS converge in 20--50
iterations to a situation where the Strehl ratio has increased from
$\sim$75\% to $\sim$99\% (a more detailed analysis  can be found  in
Section~\ref{sec:meassimu}). After the convergence, the control law,
Eq.~\eqref{eq:feedback}, gives phase updates that are negligible
compared to the shape of the wavefront corrector, $\theta_k$. However,
we run the algorithm for a total 400 iterations to make sure that no 
creeping instabilities occur.

\begin{figure}[hbtp]  \center
\includegraphics{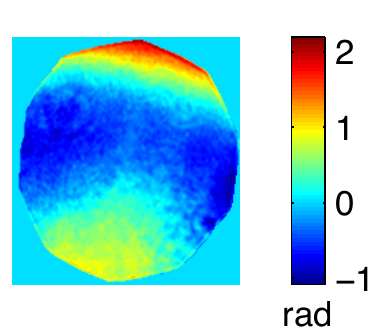}  
\includegraphics{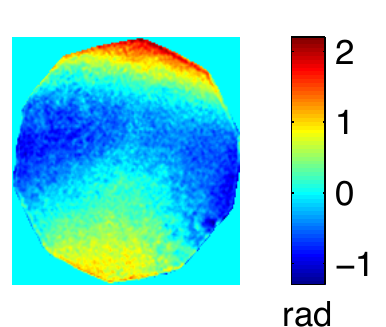} 
\includegraphics{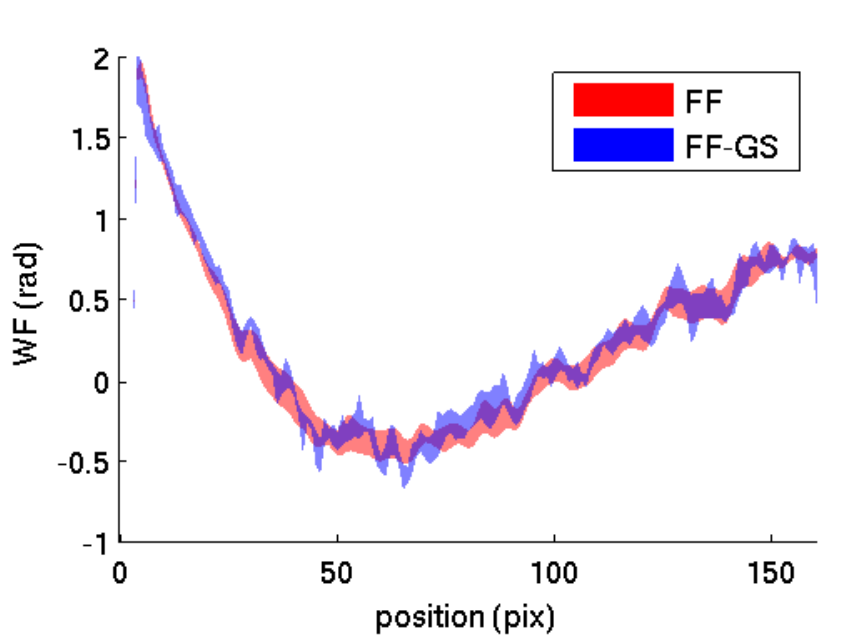} 
\caption{Top row: typical wavefront shapes ($170\times 170$~pixels) of
  the SLM after the convergence of FF and FF-GS. Bottom: radial cuts 
  through the wavefronts; the shaded area shows the range (minima and
  maxima) of five independent measurements.}
\label{fg:slmconv}
\end{figure}

Fig.~\ref{fg:slmconv} illustrates the typical wavefronts we obtained
after the convergence. Due to the applied low-pass filter, FF yields
wavefronts smoother than FF-GS; otherwise they match
well, though. The repeatability of the experiments appears reasonable:
the converged wavefront shapes have experiment-to-experiment differences at 
most $\sim$0.2--0.3~rad. The spread of the FF-GS results tends to be smaller
compared to FF, and we see that also the higher spatial frequencies are produced
in a repeatable way.

\begin{figure}[hbtp]  \center
\includegraphics[width=0.32\columnwidth]{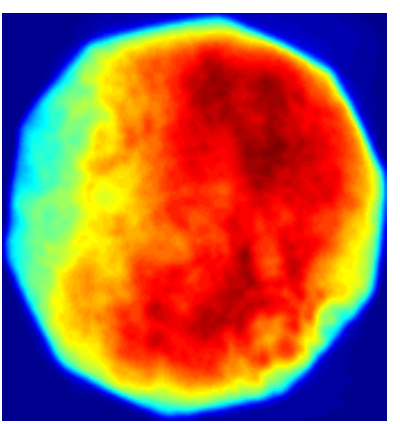}
\includegraphics[width=0.32\columnwidth]{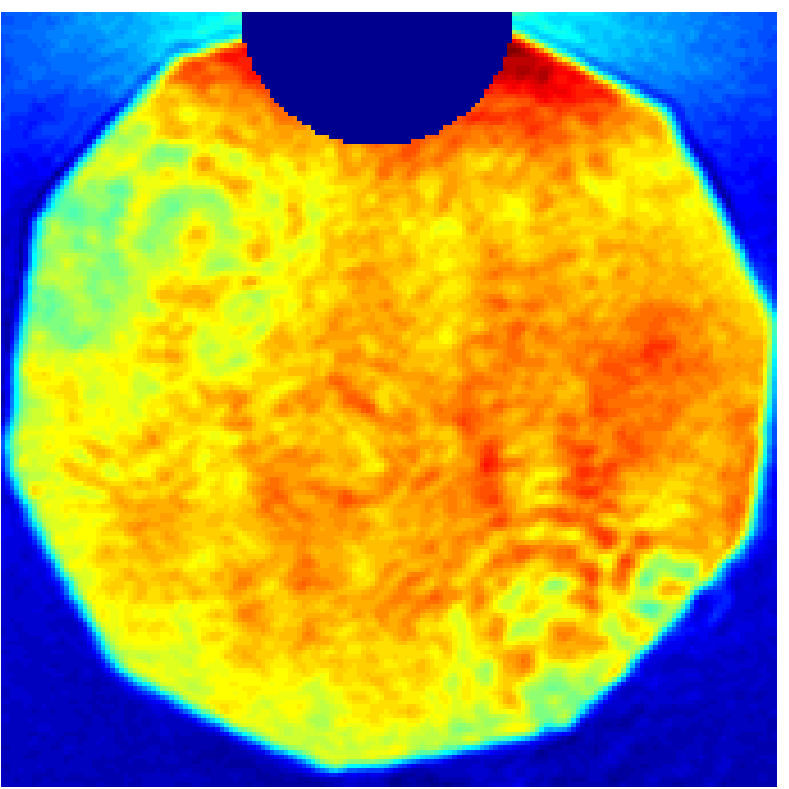}
\includegraphics[width=0.32\columnwidth]{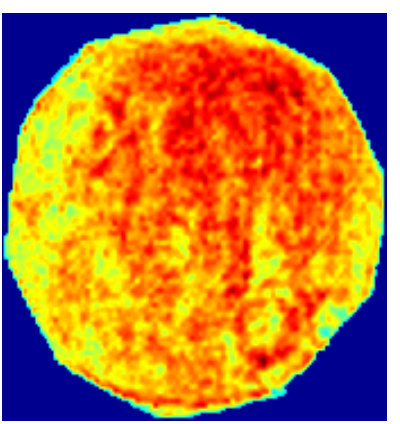}
\includegraphics{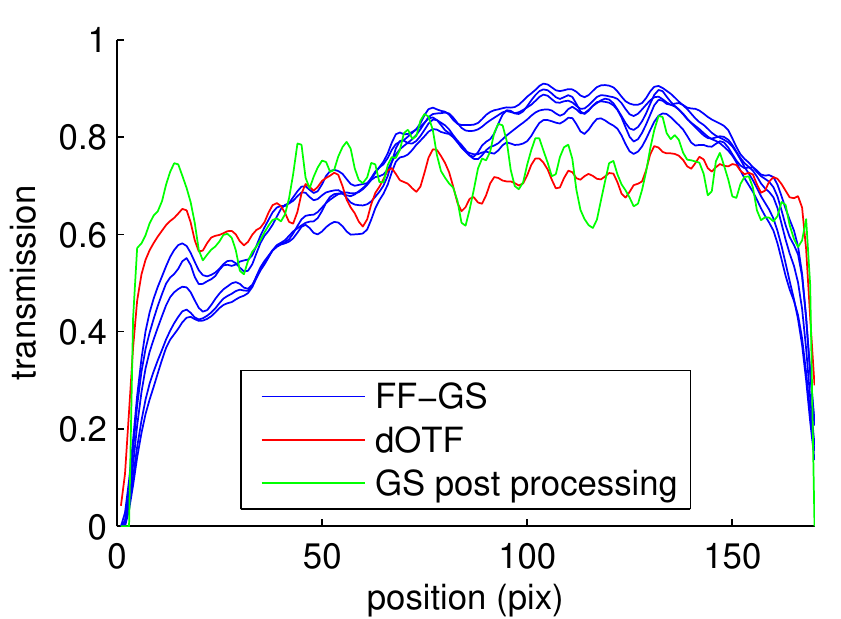}

\caption{Top row: pupil amplitudes ($170\times 170$~pixels) 
  reconstructed with different methods. Left: FF-GS. Middle: dOTF (same as
  in Fig.~\ref{fg:dotf}). Right: GS post-processing from a converged PSF.  
  Bottom: radial cuts through the pupil amplitudes; five independent
  measurements runs shown for FF-GS.}
\label{fg:pupest1}
\end{figure}

Fig.~\ref{fg:pupest1} shows the reconstructed pupil
amplitudes. The top left shows an average of $A$ following the
application of Eq.~\eqref{eq:ags} during a total of 400 FF-GS iterations
with phase updates. It can be compared with the dOTF modulus shown 
next to it, and we see that the shape of the diaphragm and
several bigger dust particles are correctly recovered. However, it is
obvious that all the finer details are lost, and the very lowest
spatial frequencies also deviate from each other. The plot below in
Fig.~\ref{fg:pupest1} shows radial cuts of five similarly obtained
pupil amplitudes, and we see that all the features in the pupil
amplitudes are nevertheless repeatedly reconstructed in the same way.

To obtain an improved reconstruction of the finer details in the pupil
amplitudes, we use the PSF that results after the FF-GS algorithm has
converged. We assume that all the remaining speckles are caused by the
amplitude aberrations, and reconstruct -- with a Gerchberg-Saxton-style
algorithm -- a pupil that would create such a pattern. This is shown
in the upper right in Fig.~\ref{fg:pupest1}, and we can see that it indeed
much better matches the dOTF reconstruction in
Fig.~\ref{fg:dotf}. Later, we use this pattern in simulations for
analysis purposes.

The differences between the independent measurement series shown here are a
combination of actual small changes in the hardware and uncertainty caused
by noise and systematic errors. It is difficult to separate those two
effects, and therefore we continue the analysis with the help of
numerical simulations.

\subsection{Comparison of measurements and simulations}
\label{sec:meassimu}

To simulate the optical setup, we assume that the algorithms correct
wavefronts shown in Fig.~\ref{fg:slmconv} with pupil amplitudes
similar to what is shown in Fig.~\ref{fg:pupest1}. We created three
study cases reflecting the variability in the converged results.

In the simulations, we consider eight different sources of errors that
needs to be modeled explicitly. They are:
\begin{enumerate}
\item SLM quantification. We use only 6 bits to control the
  wavefront. The plots shown in Fig.~\ref{fg:slmrespo} are used to
  round the simulated WF correction to what would happen in practice.

\item PSF sampling. The wavefront and the resulting PSF are sampled
  internally by a factor of two higher than what the hardware controls
  or observes. The control algorithms use re-binned PSFs, and the
  simulated wavefront correction is interpolated bilinearly from
  the reconstruction at a resolution of $170\times 170$.

\item Image noise and dynamic range. We estimate the read-out noise of
  the HDR images to be at a level of $2.2\cdot 10^{-6}$ of the image
  maximum. Gaussian random noise is added to the simulated PSFs. The
  HDR images have maximum values $\sim$$4\cdot 10^8$,
  corresponding to about 29 bits, and this is also modeled in the simulations.

\item Background level. Standard background subtraction is performed
  on the PSF images, but a small error will still
  remain. Therefore, we add a constant background level,
  $2.7\cdot 10^{-6}$ of the image maximum, to the simulated PSFs.

\item Non-perfect pupil. Instead of the perfect top-hat function, we
  use pupil amplitudes similar to what is illustrated in the top right of
  Fig.~\ref{fg:pupest1}.

\item Amplitude aberrations. We simulate the coupling of the wavefront
  and the transmission of the SLM as illustrated by
  Fig.~\ref{fg:slmrespo}.

\item Alignment errors. Although the dOTF calibration is rather accurate,
  some error could still be present in the affine transform that we use to
  map the wavefront to the SLM pixels. The simulations indicate that
  if the transform has a mismatch corresponding to a rotation larger
  than 0.4$^\circ$, FF and FF-GS would be unstable. In practice, with
  the used hardware, we saw no hints these of
  instabilities. Therefore, a rotation error of 0.4$^\circ$
  represents the maximum misregistration that the wavefront control
  algorithms are likely to experience.

\item Tip-tilt error. Internal turbulence in the optical setup causes
  frame-to-frame wavefront variations, which can be approximated
  to a degree as small shifts of the recorded images.
  We measured the difference of the
  center-of-gravity between two consecutive PSFs recorded with the HDR
  method, and it was found to be on average 0.025~pixels.
  This error cannot be taken into account by the phase-diversity
  approach, and we model its impact on the performance.
  
\end{enumerate}

\begin{figure}[hbtp]  \center
\includegraphics{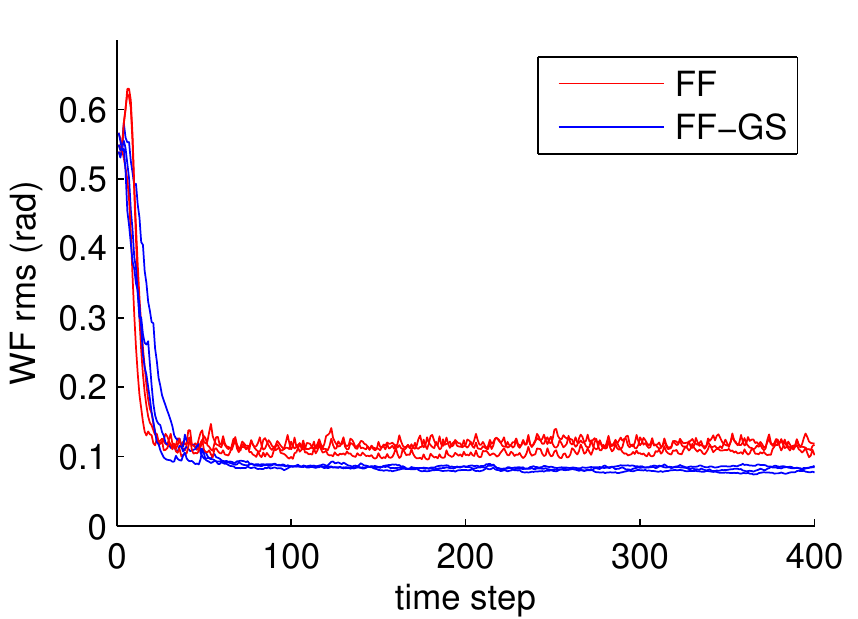} 
\includegraphics{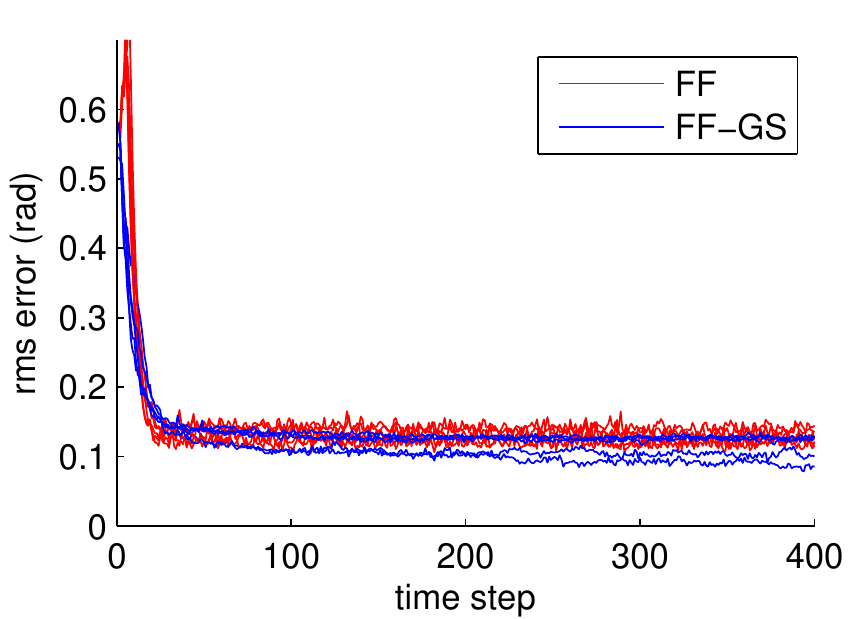} 
\caption{Tip/tilt removed residual wavefront error as a function of
  time step. Top: simulations (real value). Bottom: measurements
  (estimation from PSF images).}
\label{fg:converg1}
\end{figure}

Fig.~\ref{fg:converg1} shows the remaining wavefront
error as a function of time step. The simulation plots show the exact
error, but the measured value is estimated from the data. Here,
we have estimated the rms error from the corresponding PSF 
images only. At first, we
estimated the Strehl ratios using the method seven in
\cite{roberts2004}, and the result was converted to an rms error using
the expression $S=\exp(-\sigma^2)$. The resulting estimates are highly
sensitive to the estimation of the pupil amplitudes, which we know
only approximately (Fig.~\ref{fg:pupest1}). Thus, the y-axis in the
lower plot in Fig.~\ref{fg:converg1} is not directly comparable to the
simulation plot; alternative estimates that are more easily compared are 
shown later in this Section.

Nevertheless, the speed of the convergence is clearly seen. Both FF
and FF-GS reduce the WF rms error from $\sim$0.5~rad rms to $\sim$0.1
in $\sim$50~iterations. FF converges about 50\% faster, but it is
plagued by the overshoot at the beginning; it would require an
adaptive optimization of the low-pass filter to properly handle it.

Regarding the simulations, it is obvious that the FF-GS improves the
performance over FF: the rms error is 0.08~rad as compared to
0.12~rad. This is largely due to the smaller value of the leaky
integrator gain that we had to apply to make the FF stable.

Regarding the measurements, we can see a similar pattern, but we also
see that the FF-GS has two modes: the estimate of the residual rms
error is either $\sim$0.10~rad or $\sim$0.13~rad. The modes are
related to the finite sampling of the CCD detector. Our models do not
explicitly constrain the position of the PSF at the detector, which
means that a random sub-pixel tip/tilt component -- different between
the independent measurement series -- is left in the residual
wavefront. The algorithms converge to a state that remains stable, but
the different remaining tip/tilt components can cause significant
changes in the measured maximum intensity, and this affects our Strehl
ratio estimation process. When inspecting the re-centered PSFs
carefully, as shown later in this section, no significant differences
between the PSFs can be seen.

\begin{figure}[hbtp]  \center
\includegraphics{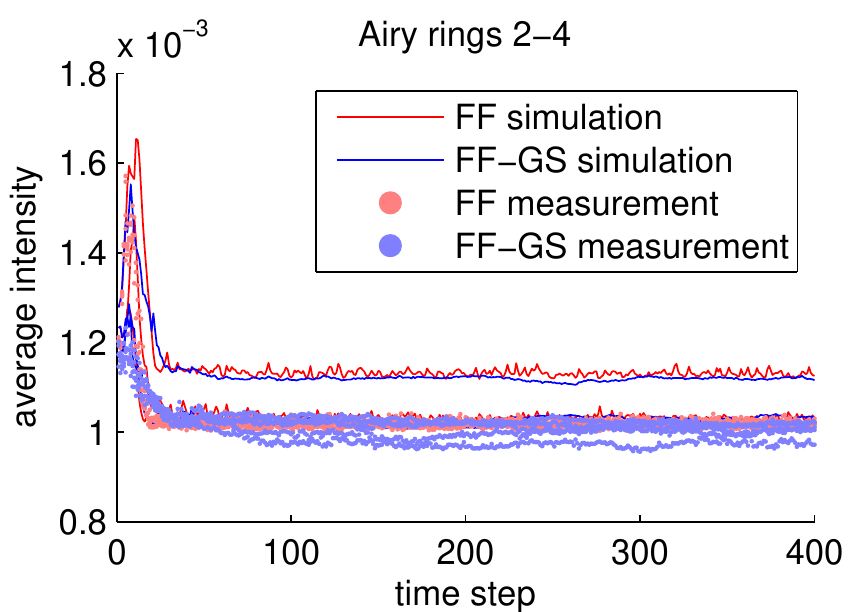} 
\includegraphics{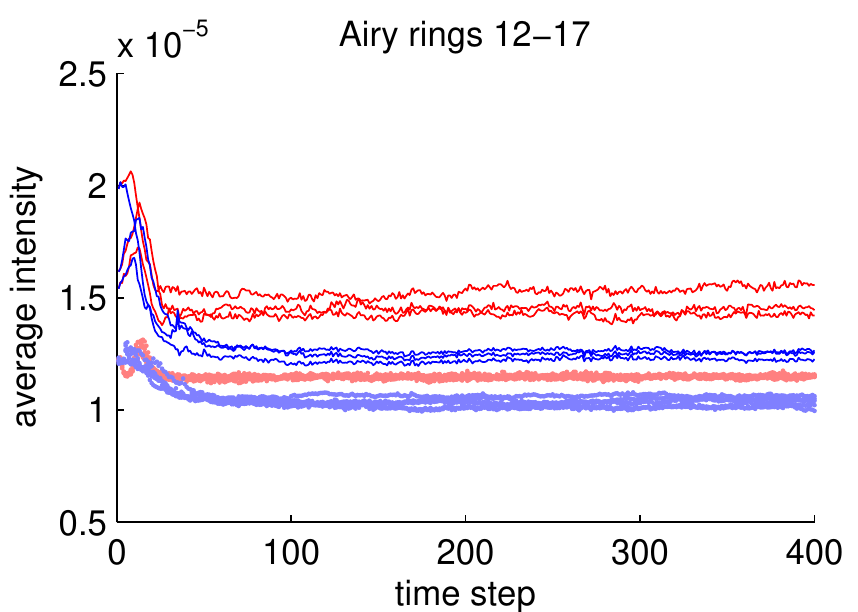} 
\includegraphics{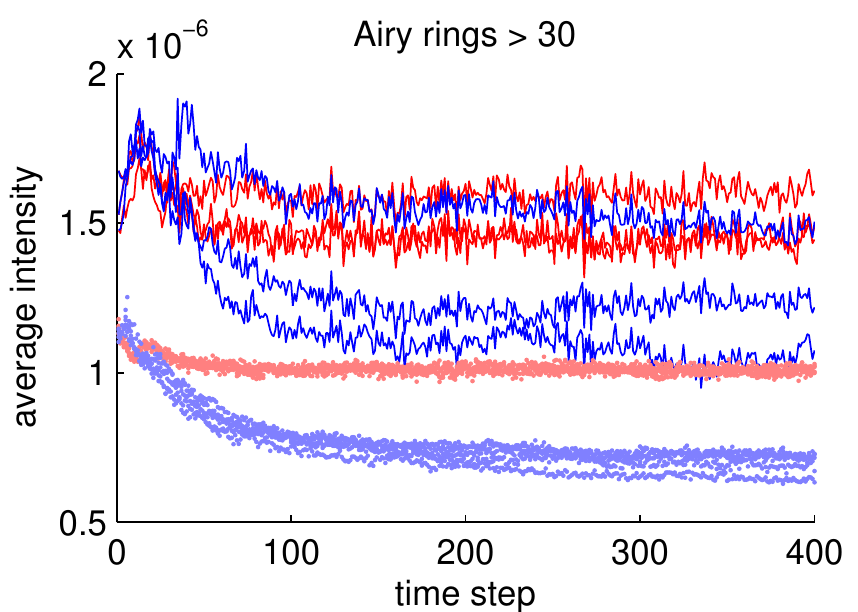} 
\caption{Average intensity at different parts of the field. Three cases are shown: the field corresponding to Airy rings 2--4, Airy rings 12--17 and Airy further than 30.}
\label{fg:psdconverg}
\end{figure}

A more detailed investigation reveals that the convergence of the wavefront correction depends on the spatial frequency -- low-frequency features are reconstructed faster. Fig.~\ref{fg:psdconverg} illustrates this by showing how an average intensity in different regions of the field changes as a function of time step. We show three different regions representing low, medium and high spatial frequencies; the locations correspond to Airy rings 2--4, 12--17 and rings further than 30. Since we consider only small wavefront aberrations, the shown intensity values are directly proportional to the average power spectral density at the matching frequency bands.

Both simulations and measurements show a similar pattern, although the absolute levels are higher in simulations due to differences in noise. At low spatial frequencies,  both FF and FF-GS peak at iterations 5--10. FF converges in total in $\sim$20 iterations, and FF-GS takes $\sim$20 iterations more, although some cases show intensity reduction even until $\sim$100 iterations. At medium spatial frequencies, the peak occurs at iteration $\sim$15, and the algorithms need in total $\sim$30 iterations to reach intensity level $\sim$6\% lower than at the beginning. FF saturates at that level, but 30 additional iterations with FF-GS reduce the intensity in total $\sim$15\% from the initial level. At high spatial frequencies, FF requires almost 50 iterations to converge to a level 15\% lower than the initial intensity (in simulations the reduction is only a few percentages due to higher noise). FF-GS, on the other hand, converges faster than FF, but still 150 iterations are needed to reduce the intensity $\sim$35\%. 
The measurements show marginally better intensity reduction, but that requires almost 300 iterations.

\begin{figure}[hbtp]  \center
\includegraphics{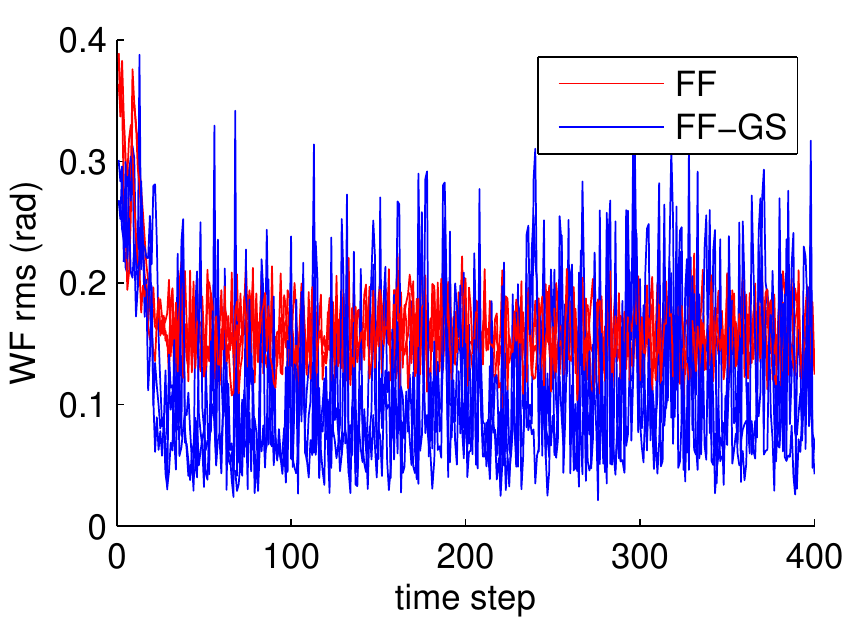} 
\includegraphics{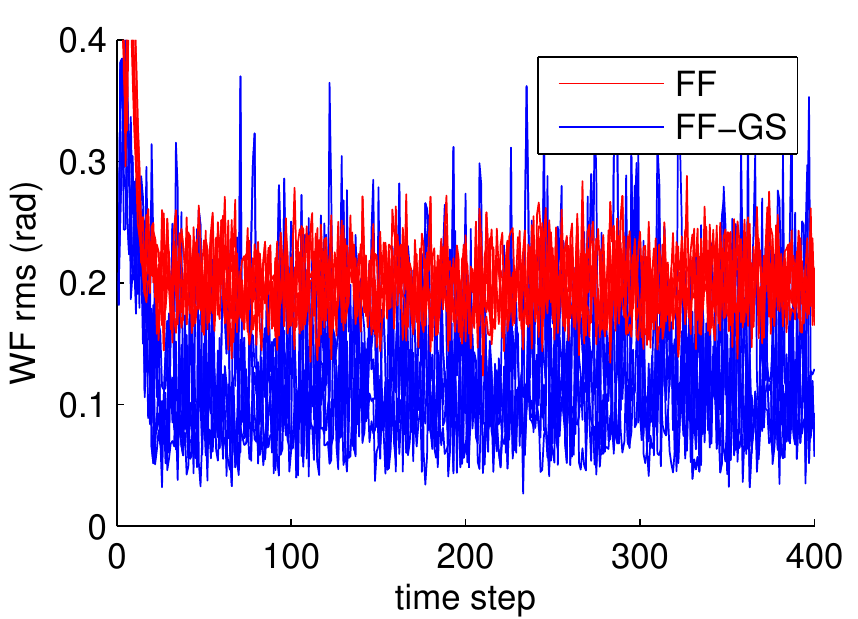} 
\caption{Residual wavefront error as a function of time step. Values
  calculated from the actual estimates used by the algorithms. Top:
  simulations. Bottom: measurements.}
\label{fg:converg3}
\end{figure}

The residual wavefront error can obviously also be estimated using
the control data that the algorithms themselves provide through
Eqs.\eqref{eq:phisol} and \eqref{eq:phigs}; the corresponding 
results are shown in Fig.~\ref{fg:converg3}.

The first striking feature is that the simulations and the
measurements produce practically identical patterns.  After the
convergence, the WF estimates of the FF algorithm have an rms error of
0.12--0.18~rad in the simulations and 0.15--0.24~rad in the
measurements. There appears to be no obvious structure in how
the error varies between consecutive iterations. Since the 
actual correction is an average over
several consecutive measurements, the actual remaining wavefront error
can be smaller than the instantaneous estimates of 0.12--0.24~rad. 
In the simulations, the error was observed to be
$\sim$0.12~rad, and we have no reason to assume the situation with the
actual hardware would be different; our estimate for the remaining
WF rms error is $\sim$0.15~rad.

With the FF-GS algorithm, the issue is slightly more complicated since
some of the WF estimates fail when the algorithm approaches the optimum. 
The reason for this -- the phase-diversity failure -- is discussed in
Section~\ref{sec:ffgspractical}.  This is seen as prominent spikes in
the plots in Fig.\ref{fg:converg3}, although most of the rms error
values are concentrated around 0.1~rad. In the simulations, the actual rms
error of the residual wavefront is $\sim$0.08~rad, and a similar value
is seen in the actual measurements.

\begin{figure}[hbtp]  \center
\includegraphics{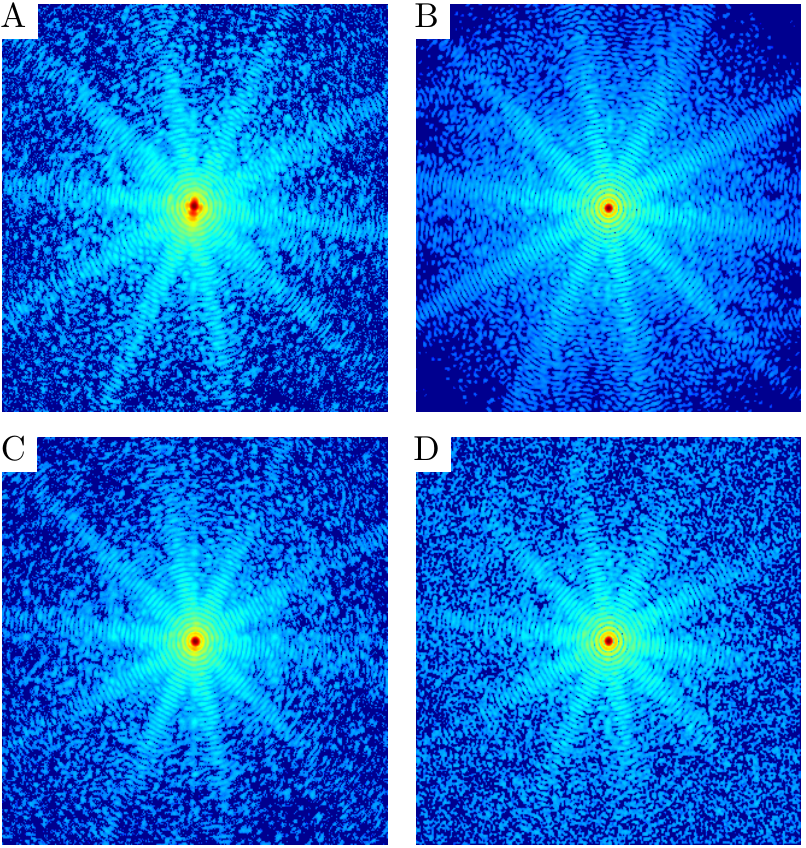}

\caption{Examples of PSF images ($320\times 320$~pixels) raised to the
  0.1 power. A) initial, measured.  B) perfect, simulated C) converged
  FF-GS, measured. D) converged FF-GS, simulated. }
\label{fg:psfsampls}
\end{figure}

Four examples of the actual PSF images are shown in Fig.~\ref{fg:psfsampls}:
\begin{enumerate}
\item the initial PSF (measured when the SLM pixels are set to zero),

\item the simulated perfect PSF resulting from the pupil amplitudes
shown in Fig.~\ref{fg:pupest1},

\item simulated PSF after the convergence of the FF-GS algorihm,

\item measured PSF after the convergence of the FF-GS algorithm.
\end{enumerate}

All the PSFs have a similar, star pattern with ten radial beams
gradually fading towards the edges of the images. These are caused by
the blades of the diaphragm, whose shape is shown in
Figs.~\ref{fg:dotf} and \ref{fg:pupest1}.

The initial PSF corresponds to a wavefront like
in Fig.~\ref{fg:slmconv}: a clearly deformed core, but still easily
recognizable Airy rings 3--20. 

The simulated, noiseless and aberration-free PSF shows the 
speckles that we expect to remain
due to the non-flat pupil amplitudes. The dust, dirt and also the
inhomogeneities of the SLM create a significant transmission
distortion dominated by high spatial frequencies. This causes the halo
of irregularities on top of the pattern of the perfect diffraction
rings. In addition, we can see a few stronger speckles and speckle
groups at a distance of approximately Airy rings 12--18. These can
be attributed to the larger dust particles also clearly visible in the
FF-GS estimated pupil amplitudes in Fig.~\ref{fg:pupest1}.

When comparing the measured and simulated PSFs after the FF-GS
algorithm has converged, we find no significant differences. Both
PSFs have a regular core, which appears to match exactly the
perfect PSF up to the 4th diffraction ring. At least 26 diffraction
rings are, at least partially, visible. A comparison with the perfect PSF
shows that several strong speckles can be identified in all the
images, but the halo after the 14th diffraction ring outside the
star-like beams, close to the detection limit of the camera, is
dominated by speckles with no obvious structure.

\begin{figure}[hbtp]  \center
\includegraphics{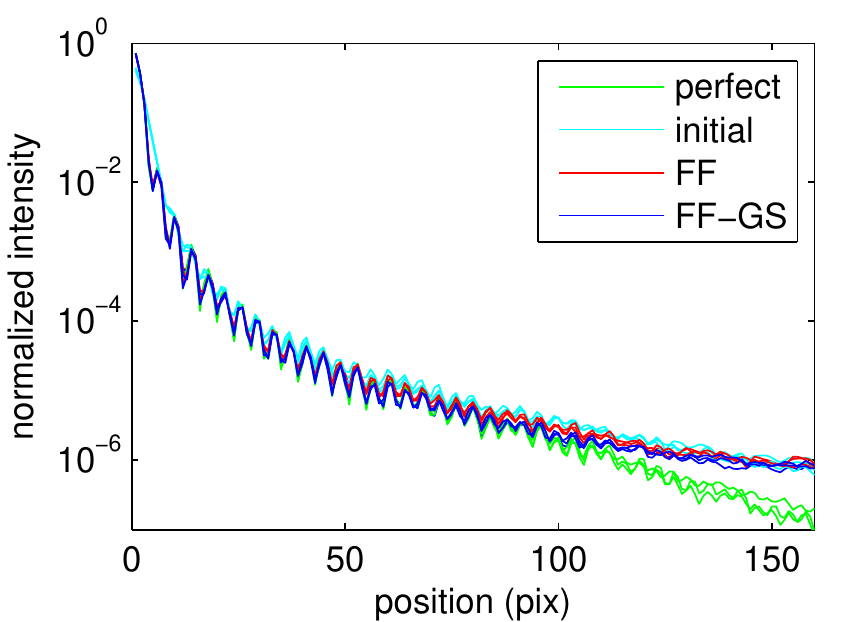} 
\includegraphics{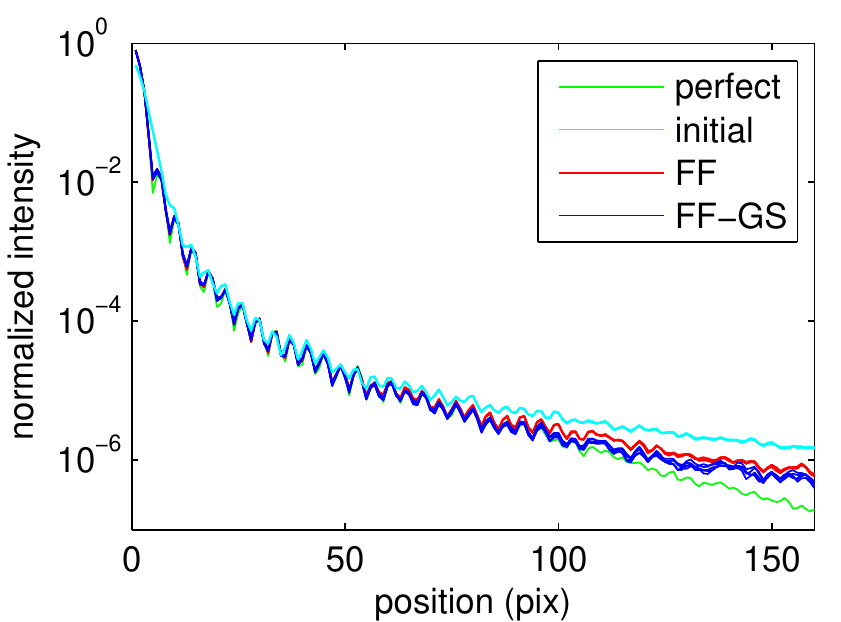} 
\caption{Averaged radial profiles of PSF images. Upper: simulated, the
  three study cases are shown. Lower: measured, results from five
  indepedent runs are shown; the perfect PSF is identical to the one in
   the upper plot.}
\label{fg:psfradprofs}
\end{figure}

A more detailed comparison can be obtained by inspecting the radially
averaged profiles of the PSFs. Before taking the radial average,
we shift, using Fourier transforms, the PSFs to have the center-of-gravity
at the location of the perfect PSF. The results are shown in
Fig.~\ref{fg:psfradprofs}.

The profiles show that both the FF and FF-GS algorithms, in both the
simulated and measured cases, converge to a situation very close to
the perfect simulated PSF; no significant differences are seen up to
the first 13 (simulated) or 20 (measured) diffraction rings. After
this, we can see that the performance of both algorithms slowly
deviates from the perfect PSF, the intensity being 
a factor of $\sim$5 (simulated) or $\sim$2--3 (measured) higher at borders.
At the distances corresponding to diffraction rings 20 and higher,
FF-GS is typically $\sim$20--30\% better in reducing the intensity as 
compared to FF.

In total we can recognize at least 30 diffraction rings before the
speckle noise makes the PSF structure too blurry to observe any
structure. Nevertheless, compared to the initial PSF, both algorithms
reduce the intensity of scattered light throughout the whole used field, 
although, in the simulated case, the difference is not significant
after the 34th diffraction ring. In the measured case, on
the other hand, the light intensity is reduced by a factor of
$\sim$2--3 also at the edge of the recorded image.
This difference between the simulations and measurements 
is due to a combined effect of differences in actual noise levels, 
wavefronts and pupil transmission.

\subsection{Error budget}
\label{sec:errbud}

Finally, we show an error budget that illustrates the impact of the
different error sources in the optical setup.

In the ideal case, we have no noise and a perfectly circular pupil that
is -- in the case of FF -- exactly known. The perfect case also uses
exactly the same imaging model in both the WF reconstruction and when
simulating the PSF images: a zero-padded FFT with a wavefront modeled
at a resolution of $170\times 170$.

We sequentially simulate each of the error sources listed in
Section~\ref{sec:meassimu}. The resulting rms errors in the converged
wavefronts are listed in Table~\ref{tb:errbud}.


\newcommand{\tbspa}{$\hspace{0.1cm}$}
\begin{table}[hbtp] \begin{center}
\caption{Error budget}
\label{tb:errbud}
\begin{tabular}{lccc}
  \hline
  &  FF$^*$ & & FF-GS$^*$ \\
  \hline
  0. No errors             &  0.03 $\pm$ 0.01 &\tbspa &  0.00 $\pm$ 0.00 \\
  1. SLM quantification    &  0.04 $\pm$ 0.01 &\tbspa &  0.02 $\pm$ 0.00 \\
  2. PSF sampling 2x       &  0.08 $\pm$ 0.01 & &  0.01 $\pm$ 0.00 \\ 
  3. Image noise           &  0.05 $\pm$ 0.01 & &  0.05 $\pm$ 0.00 \\
  4. Background level      &  0.04 $\pm$ 0.01 & &  0.01 $\pm$ 0.00 \\
  5. Non-perfect pupil     &  0.11 $\pm$ 0.00 & &  0.02 $\pm$ 0.01 \\
  6. Amplitude aberrations &  0.12 $\pm$ 0.01 & &  0.04 $\pm$ 0.01 \\
  7. Alignment errors      &  0.08 $\pm$ 0.01 & &  0.01 $\pm$ 0.00 \\
  8. TT instability        &  0.03 $\pm$ 0.01 & &  0.04 $\pm$ 0.01 \\
  9. All errors            &  0.12 $\pm$ 0.01 & &  0.08 $\pm$ 0.00 \\
  \hline
\end{tabular}\\
$^*$The residual WF rms errors (rad) at spatial frequencies falling
within the used images.
\end{center} \end{table}

In theory, both algorithms should reach zero wavefront error in
the perfect case. However, in the case of FF, we still have to use
numerical regularization to maintain stability, and this compromises the
performance in the error-free case. This could be improved by optimizing
the codes, but it is not done here; the codes are optimized for the
performance with all the error sources present.

The most severe error source for the FF algorithm, as expected, is
indeed the amplitude aberrations: instead of the ideal rms error of
0.03~rad, we are limited to an error of 0.11~rad. Similar errors are
also seen if the imaging model does not exactly match the actual
hardware; this was tested when simulating the wavefront and PSF with
double sampling (case 2 in Table~\ref{tb:errbud}); the double sampling was also
used in the misalignment simulation. The different error sources are
coupled, so they do not add up quadratically. In the presence of all
the error sources, we end up having a residual WF error of
$\sim$0.12~rad.

With the FF-GS algorithm, we can radically reduce the problems of the
unknown pupil aberrations. The transmission we used in simulations,
however, had significant fluctuations creating speckles similar to
what the wavefront aberrations do. Therefore, the wavefront
reconstruction problem is difficult to make unambiguous, and
we saw a small residual rms error of 0.02~rad.

The FF-GS is limited by the combined effect of read-out noise
(0.05~rad), the fact that the SLM couples the transmission and phase
change (0.04~rad) and the TT instability (0.04~rad). All the error
sources add up quadratically, which indicates that they are largely independent.

When comparing the FF and FF-GS, we see that a significant improvement
can be obtained with the FF-GS algorithm; the residual wavefront rms
error is reduced from 0.12~rad to 0.08~rad. However, the method is
more sensitive to uncertainties and noise: the tip-tilt jitter in our
hardware has no influence on the FF while being a major error source
with the FF-GS algorithm.

\section{Conclusions and discussion}
\label{sec:conclusions}

We have demonstrated the performance of two numerically efficient
focal-plane wavefront sensing algorithms: the Fast \& Furious and its
extension Fast \& Furious Gerchberg-Saxton.

Both algorithms do an excellent job in calibrating static aberrations
in an adaptive or active optics system: we demonstrated an increase in
the Strehl ratio from $\sim$0.75 to 0.98--0.99 with our optical setup.

Although the FF-GS algorithm is more prone to noise,
we observed a clear improvement. With our hardware -- a
high-resolution spatial light modulator as the wavefront corrector
-- we estimate the remaining residual wavefront rms error to be
$\sim$0.15~rad with FF and $\sim$0.10~rad with FF-GS. The difference
occurs mostly at spatial frequencies corresponding to the 20th and
further Airy rings.

Simulations with error sources comparable to our hardware show
very similar results. This increases our  confidence that the estimated
performance indicators are reliable, and the simulated error budget
also confirms the unknown amplitude aberrations as the main limitation
of the FF algorithm in the considered framework.

To our knowledge, this is the first time that such focal-plane sensing
methods have been demonstrated with $\sim$30~000 degrees of freedom -- and
in the case of FF-GS, with twice the number of free parameters
to estimate the pupil amplitudes.

The sampling at the detector was such that the controlled wavefront of
$170\times 170$ pixels would have been enough to correct all
spatial frequencies inside an image of $640\times 640$
pixels. However, as we recorded only an image of $320\times 320$
pixels, we had no direct observations of the 
higher controlled spatial frequencies. Simulations indicate that this
resulted in a small amount of light being scattered outside the
recorded field, but this amount was too small to be easily detected in
our optical setup.

We put no particular effort into optimizing the codes; all the software
was implemented in Matlab, and it was run on a standard Windows PC.
Still, the required computation time was negligible compared to the
time of $\sim$15~s we needed to collect data for a single HDR
image. We implemented the FF algorithm with two $640\times 640$ FFTs per
iteration step (one FFT transferring the phase-diversity
information into the focal plane could likely be replaced by a
convolution, as explained in \cite{keller2012spie}). 
Our FF-GS implementation used 8 FFTs per iteration,
and that could also potentially be optimized.

As with all focal-plane wavefront sensing techniques, the
algorithms work best if a monochromatic light source is available.
With a chromatic light source having a sufficiently small bandwidth,
perhaps $\sim$10\%, the algorithms would still work, but only with a
limited corrected field. With special chromatic optics (such as in
\cite{guyon2010}) or an integral field unit, it is
potentially possible to use the algorithms with even wider bandwidth.

Currently, we have only demonstrated a case where an unobstructed PSF
is detected, and the wavefront is driven to be flat. To make the
algorithms more interesting for astronomical applications in the
extreme adaptive optics or ultra-high contrast imaging, a few
extensions would be necessary.

First, we should consider how coronagraphs and diffraction suppression
optics will affect the techniques. In practice, this would mean that
the core of the PSF would not be detected, and we would need to
consider also the moduli in a part of the focal-plane field as free
parameters. 

Second, instead of flattening the wavefront, we should optimize the
contrast at a certain part of the field. This would mean calculating
a wavefront shape that, in the same way as in \cite{malbet1995, borde2006, giveon2007oe}, minimizes
the light in certain regions of the field at the cost of increasing it
in other parts; the updated algorithm should then drive the wavefront
to this desired shape. A similar problem is faced, if phase plates are
used to create diffraction suppression, for instance as in \cite{codona2004}.
Also in such a case, it is necessary to drive the wavefront to a particular 
shape that is far from flat.

Another, potentially interesting application is a real-time
application, for instance as a high-order, second-stage sensor in an
adaptive optics system. The computational load is manageable, and a
successful system would greatly simplify the hardware design compared
to a conventional AO approach. However, issues such as the requirement
for small aberrations, chromaticity, temporal lag in the phase diversity and 
the limited dynamic range of the camera -- and therefore photon noise --
are major challenges.

\bibliographystyle{osajnl}

\end{document}